\documentclass[twocolumn,preprint2]{aastex6}

\usepackage{ulem}
\usepackage{xcolor}
\usepackage{amsmath}
\usepackage{cases}
\usepackage[T1]{fontenc}

\begin{document}
\title[]{Supernova luminosity powered by magnetar-disk system}

\author{W. L. Lin\altaffilmark{1}, X. F. Wang\altaffilmark{1,2,3}, L. J. Wang\altaffilmark{4}, Z. G. Dai\altaffilmark{5}}
\altaffiltext{1}{Physics Department and Tsinghua Center for Astrophysics (THCA), Tsinghua University, Beijing 100084, China; \\
linwl@mail.tsinghua.edu.cn; wang\_xf@mail.tsinghua.edu.cn}
\altaffiltext{2}{Beijing Planetarium, Beijing Academy of Science and Technology, Beijing 100044, China}
\altaffiltext{3}{Purple Mountain Observatory, Chinese Academy of Science, Nanjing 210008, China}
\altaffiltext{4}{Astroparticle Physics, Institute of High Energy Physics, Chinese Academy of Sciences, Beijing 100049, China}
\altaffiltext{5}{CAS Key Laboratory for Research in Galaxies and Cosmology, Department of Astronomy, University of Science and Technology of China, Hefei 230026, China}

\begin{abstract}
Magnetars are one of the potential power sources for some energetic supernova explosions such as type I superluminous supernovae (SLSNe I) and broad-lined type Ic supernovae (SNe Ic-BL). In order to explore the possible link between these two subclasses of supernovae (SNe), we study the effect of fallback accretion disk on magnetar evolution and magnetar-powered SNe. In this scenario, the interaction between a magnetar and a fallback accretion disk would accelerate the spin of the magnetar in the accretion regime but could result in substantial spin-down of the magnetars in the propeller regime. Thus, the initial rotation of the magnetar plays a less significant role in the spin evolution. Such a magnetar-disk interaction scenario can explain well the light curves of both SNe Ic-BL and SLSNe I, for which the observed differences are sensitive to the initial magnetic field of the magnetar and the fallback mass and timescale for the disk. Compared to the magnetars powering the SNe Ic-BL, those accounting for more luminous SNe usually maintain faster rotation and have relatively lower effective magnetic fields around peak time. In addition, the association between SLSNe I and long gamma-ray bursts, if observed in the future, could be explained in the context of magnetar-disk system.

\end{abstract}


\section{Introduction}

Type I superluminous supernovae (SLSNe I; \citealp[e.g.,][]{2012Sci...337..927G, 2019ARA&A..57..305G, 2019NatAs...3..697I}) are a newly-discovered type of the most luminous supernovae (SNe) whose early-time spectra are dominated by O\,\textsc{ii} absorption complexes and blue continua indicating high photospheric temperature \citep[e.g.,][]{2011Natur.474..487Q, 2018ApJ...855....2Q}. Although SLSNe~I exhibit distinct early-time light curves and spectral features compared to normal and broad-lined SNe Ic (SNe Ic-BL), the similarity in their late spectra \citep[e.g.,][]{2010ApJ...724L..16P, 2019ApJ...872...90B, 2020MNRAS.497..318L} implies an intrinsic link between these two subclasses of SNe with both hydrogen and helium envelope stripped before explosion. A systematic comparison study conducted by \citet{2017ApJ...845...85L} reveals that the absorption features, such as the widths and average velocities of Fe\,\textsc{ii} $\lambda5169$, are similar in the mean post-peak spectra of SLSNe I and SNe Ic-BL, while normal SNe Ic usually exhibit narrower absorption lines with a lower blueshift velocity. Similarities between SLSNe I and SNe Ic/Ic-BL can be observed in their nebular-phase spectra; especially in the iron-dominated wavelength range of $4000-5500$ \AA, SLSNe I and SNe Ic-BL have more properties in common as compared to normal SNe Ic \citep{2019ApJ...871..102N}. 

Despite several models have been proposed so far for energetic core collapse SNe \citep[e.g.,][and references therein]{2019ARA&A..57..305G, 2019RAA....19...63W}, the spin-down of magnetar, i.e. strongly magnetized neutron star (NS), has been invoked as a promising mechanism to power SLSNe I and SNe Ic-BL \citep[e.g.,][]{2010ApJ...717..245K, 2010ApJ...719L.204W, 2013ApJ...770..128I, 2017ApJ...851...54W, 2017ApJ...837..128W}. Moreover, both SLSNe~I and SNe~Ic-BL tend to occur in faint dwarf hosts with low metallicity (e.g., \citealp{2014ApJ...787..138L, 2016ApJ...830...13P, 2018MNRAS.473.1258S, 2020ApJ...892..153M}), indicating that they are associated with metal-poor massive progenitor stars. During the evolution of such progenitor stars, stellar wind might be reduced and sufficient angular momentum can be sustained in aid of the formation of fast spinning magnetars. Although most SNe~Ic are found in higher metallicity environments \citep{2020ApJ...892..153M} and prefer radioactive decay of $^{56}$Ni as the main power source, a small portion of them exhibit engine-powered properties \citep[e.g.,][]{2015Natur.523..189G, 2016ApJ...828L..18N, 2018AA...609A.106T, 2019AA...621A..64T}. In the isolated magnetar-powered scenario, the magnetars for SLSNe I possess an initial spin period $P_{\mathrm{NS},0}\approx1-10$ ms and surface magnetic field $B_\mathrm{NS}\sim10^{12}-10^{14}$ G, while those with $P_{\mathrm{NS},0}\gtrsim10$ ms and $B_\mathrm{NS}>10^{14}$ G are expected to power SNe Ic/Ic-BL. \citet{2020ApJ...903L..24L} proposed that the above $B_\mathrm{NS}-P_{\mathrm{NS},0}$ correlation is consistent with the relation of $B_\mathrm{NS}\propto P_\mathrm{eq}^{7/6}$ expected in an equilibrium state reached during the interaction between a magnetar and an accretion disk (e.g., \citealp{2011ApJ...736..108P}).

In this paper, we study the evolution of a magnetar surrounded by a fallback accretion disk and explore the possibility that both SLSNe I and SNe Ic-BL can be produced in such a magnetar-disk scenario. In Section \ref{Sec: model}, we develop a magnetar-disk model to study the effect of fallback accretion on the magnetar and the SNe powered by such a magnetar-disk system. In Section \ref{Sec: results}, we study the effect of initial properties of the magnetar-disk system on the luminosity evolution of SNe. A brief conclusion is presented in Section \ref{Sec: conclusion}.

\section{Model description}
\label{Sec: model}
\subsection{Evolution of a magnetar with a disk}
\label{SubSec: model_NS} 

A rapidly rotating magnetar might be born in SN explosion, and a portion of stellar debris could fall back to circularize into a disk around the magnetar with an accretion rate greatly exceeding the Eddington limit ($\dot{M}_\mathrm{Edd}$). The highly super-Eddington accretion disk is expected to be geometrically thick and probably advective \citep[e.g.,][]{1998MNRAS.297..739B}, which likely drives large-scale outflows within the time range of our interest ($\gtrsim1000$ s since SN explosion) \citep[see][and references therein]{2013ApJ...772...30D}. Assuming the accretion rate at the outer radius of the disk to be the fallback mass rate \citep[e.g.,][]{1988Natur.333..644M, 2018ApJ...857...95M}, we have
\begin{equation}
\dot{M}_\mathrm{D,out}=\dot{M}_\mathrm{fb}=\frac{2M_\mathrm{fb}}{3t_\mathrm{fb}}\left(1+t/t_\mathrm{fb}\right)^{-5/3},
\label{eq: Md_fb}
\end{equation}
where $M_\mathrm{fb}$ is the total fallback mass available for the disk, $t_\mathrm{fb}$ is the fallback timescale. Due to the presence of accompanied outflows, only a fraction ($\eta$) of the accretion rate would reach the inner disk radius, i.e. 
\begin{equation}
\dot{M}_\mathrm{D,in}=\eta \dot{M}_\mathrm{D,out}.
\label{eq: Md_in}
\end{equation}
Considered the effects of advection process and mass outflows, \citet{2019MNRAS.484..687M} found $\eta\gtrsim 0.6$ ($\eta\gtrsim 0.4$) when the disk outflow is powered by half (all) of the viscously dissipated energy. Their numerical simulations also show that $\eta$ tends to approach the minimum as the initial accretion rate increases from $1$ to $\sim1000$ $\dot{M}_\mathrm{Edd}$, which is far exceeded in all cases we consider (see Section \ref{Sec: results}). Here we ignore the possible effect of chemical composition of the disk, and take $\eta=0.5$.

\begin{figure}[t]
\includegraphics[angle=0,width=0.5\textwidth]{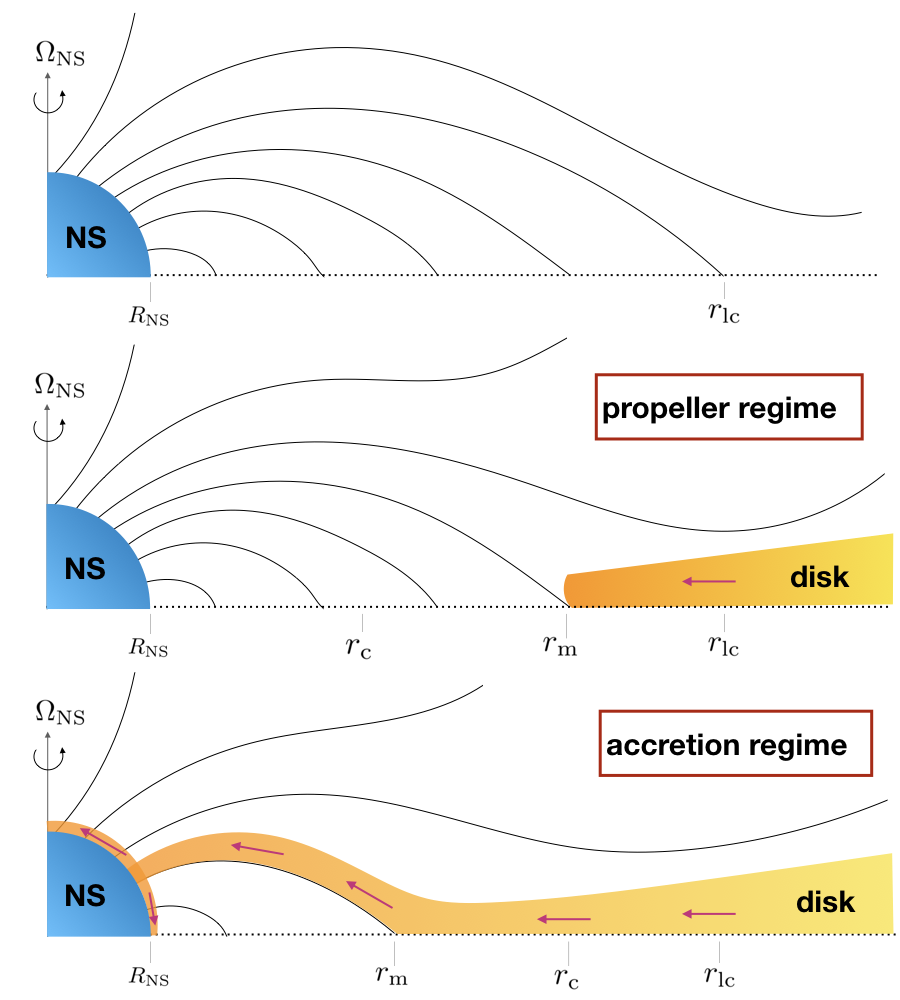}
\caption{Schematic pictures for (1) an isolated magnetar (top panel), and (2) a magnetar surrounded by a fallback accretion disk during the propeller regime ($r_\mathrm{c}<r_\mathrm{m}$, middle panel) and accretion regime ($r_\mathrm{m}<r_\mathrm{c}$, bottom panel), respectively.}
\label{fig: pic}
\end{figure}

The evolution stage of this magnetar-disk system depends on the relative position of co-rotation radius ($r_\mathrm{c}$), light cylinder radius ($r_\mathrm{lc}$), magnetospheric radius ($r_\mathrm{m}$), which are related to the gravitational mass ($M_\mathrm{NS}$), the radius ($R_\mathrm{NS}$), the spin period ($P_\mathrm{NS}$) and the surface magnetic field strength ($B_\mathrm{NS}$) of the central magnetar as well as the accretion rate of the disk. We assume the magnetospheric radius to be the maximum between Alfv\'en radius ($r_\mathrm{A}$) and the radius of the magnetar ($R_\mathrm{NS}$),
\begin{equation}
r_\mathrm{m}=\max(r_\mathrm{A}, R_\mathrm{NS}).
\label{eq: rm}
\end{equation} 
Alfv\'en radius, where the radial inflow of the disk materials is blocked by the magnetic barrier of the central magnetar, is given by\footnote{The geometrical thickness of the disk could affect Alfv\'en radius by a factor of $\sim1$ \citep[e.g.,][]{2019A&A...626A..18C}.}
\begin{equation}
r_\mathrm{A}=(GM_\mathrm{NS})^{-1/7}\mu_\mathrm{NS}^{4/7}\dot{M}_\mathrm{D,in}^{-2/7}
\label{eq: ra}
\end{equation}
where $G$ is the gravitational constant and $\mu_\mathrm{NS}=B_\mathrm{NS}R_\mathrm{NS}^3$ is the magnetic dipole moment of the magnetar.
Co-rotation radius is defined as
\begin{equation}
r_\mathrm{c}=(GM_\mathrm{NS}/\Omega_\mathrm{NS}^2)^{1/3},
\label{eq: rc}
\end{equation}
where the inflowing matter revolves at the angular frequency of the magnetar ($\Omega_\mathrm{NS}=2\pi/P_\mathrm{NS}$). The light cylinder radius of the magnetar is
\begin{equation}
r_\mathrm{lc}=c/\Omega_\mathrm{NS},
\label{eq: rlc}
\end{equation}
where $c$ is the light speed.

If the disk penetrates the light cylinder of the magnetar ($r_\mathrm{m}<r_\mathrm{lc}$) and cuts open part of closed magnetic field lines, the magnetic dipole radiation wind from the magnetar would be enhanced. Thus, the magnetic dipole torque can be expressed as \citep{2016ApJ...822...33P, 2018ApJ...857...95M}
\begin{equation}
N_\mathrm{dip}=-\mu_\mathrm{NS}^2\Omega_\mathrm{NS}^3/(6c^3)\cdot
\begin{cases}
r_\mathrm{lc}^2/r_\mathrm{m}^2, & r_\mathrm{m}< r_\mathrm{lc}\\
1. &r_\mathrm{m}> r_\mathrm{lc} 
\label{eq: N_dip}
\end{cases} 
\end{equation}
and the effective magnetic field strength for the dipole radiation is $B_\mathrm{NS,eff}=B_\mathrm{NS}\cdot\max(1,r_\mathrm{lc}/r_\mathrm{m})$. 

If $r_\mathrm{m}<r_\mathrm{c}$, the disk materials at inner radius revolve faster than the magnetar and tend to be magnetically channelled towards the magnetar, resulting in the spin-up of the magnetar (accretion regime); conversely if $r_\mathrm{c}<r_\mathrm{m}$, the angular momentum of the magnetar is transferred to the inner disk (propeller regime; e.g., \citealp{1975A&A....39..185I}; see also Figure \ref{fig: pic})\footnote{In the propeller regime when $r_\mathrm{m}\sim r_\mathrm{c}$, quasi-periodic events of accretion might occur, since the insufficiently-accelerated propeller matter could pile up in the inner disk until an event of accretion onto the magnetar surface is triggered and empties the mass accumulated during the propeller state \citep{2010MNRAS.406.1208D}. In that case, the effective transition between the accretion and propeller phases might be slightly shifted to $r_\mathrm{m}/r_\mathrm{c}\gtrsim1$.}. In the propeller state, inner disk matter could be accelerated to a super-Keplerian velocity and then form a centrifugally driven outflow. It could hinder the in-falling of outer disk matter, resulting in a decrease in the accretion rate. However, the propeller outflow can be decelerated in turn, and hence a fraction of it might accumulate in the inner disk supplying extra accretion mass. Hence, we caution that the actual evolution of accretion rate might deviate from Equation \ref{eq: Md_in}.

The accretion torque that exerts on the magnetar can be given by \citep{2005ApJ...623L..41E, 2011ApJ...736..108P}
\begin{equation}
N_\mathrm{acc}= (1- \omega)\dot{M}_\mathrm{D,in}r_\mathrm{m}^{2}\Omega_\mathrm{K,m}=(1- \omega)\dot{M}_\mathrm{D,in}(GM_\mathrm{NS}r_\mathrm{m})^{1/2},
\label{eq: N_acc_m}
\end{equation}
where $\omega\equiv\Omega_\mathrm{NS}/\Omega_\mathrm{K,m}=(r_\mathrm{m}/r_\mathrm{c})^{3/2}$ is defined as the fastness parameter, and $\Omega_\mathrm{K,m}=(GM_\mathrm{NS}/r_\mathrm{m}^3)^{1/2}$ is the local Keplerian angular frequency at $r_\mathrm{m}$. If the magnetar-disk interaction dominates over the magnetic dipole radiation, this system tends to evolve towards $r_\mathrm{c}=r_\mathrm{m}$. When $r_\mathrm{m}$ equals $r_\mathrm{c}$, the magnetar reaches an equilibrium spin period \citep[e.g.,][]{2011ApJ...736..108P}
\begin{equation}
P_\mathrm{eq}=2\pi(GM_\mathrm{NS})^{-5/7}\mu_\mathrm{NS}^{6/7}\dot{M}_\mathrm{D,in}^{-3/7}.
\label{eq: Peq}
\end{equation}

Considering both effects of dipole and accretion torques, the angular momentum of the magnetar evolves as
\begin{eqnarray}
I_\mathrm{NS}\frac{\mathrm{d}\Omega_\mathrm{NS}}{\mathrm{d}t}=N_\mathrm{dip}+N_\mathrm{acc},
\label{eq: J}
\end{eqnarray}
where the moment of inertia for the magnetar is estimated as \citep{2005ApJ...629..979L}
\begin{equation}
I_\mathrm{NS}=0.237M_\mathrm{NS}R_\mathrm{NS}^2\left[1+4.2\frac{M_\mathrm{NS}/R}{M_\odot/\mathrm{km}}+90\left(\frac{M_\mathrm{NS}/R}{M_\odot/\mathrm{km}}\right)^{4}\right].
\end{equation}

The mass rate accreted onto the magnetar surface can be estimated as \citep{2011ApJ...736..108P, 2018ApJ...857...95M}
\begin{equation}
\dot{M}_\mathrm{acc}=\begin{cases}
\dot{M}_\mathrm{D,in}, &r_\mathrm{m}< r_\mathrm{c}\\
0. & r_\mathrm{m}> r_\mathrm{c}
\end{cases} 
\end{equation}
Accordingly, the baryon mass of the magnetar with initial mass of $M_\mathrm{NS,b,0}$ is $M_\mathrm{NS,b}=M_\mathrm{NS,b,0}+\int \dot{M}_\mathrm{acc} \mathrm{d}t$, and the corresponding gravitational mass can be obtained by solving $M_\mathrm{NS,b}=M_\mathrm{NS}(1+0.075M_\mathrm{NS})$ \citep{1996ApJ...457..834T}. Pile-up of the accreted matter on the surface of magnetar could cause decay of the magnetic field as \citep{1989Natur.342..656S, 1986ApJ...305..235T, 2013ApJ...775..124F}
\begin{equation}
B_\mathrm{NS}=B_\mathrm{NS,0}/(1+M_\mathrm{acc}/M_\mathrm{c}),
\label{eq: B_NS}
\end{equation}
where $B_\mathrm{NS,0}$ is the initial magnetic field. As for the uncertain characteristic mass $M_\mathrm{c}$, we follow \citet{2021ApJ...907...87L} to adopt $M_\mathrm{c}=10^{-3}M_\odot$. The magnetic field will re-diffuse to the surface of NS due to Ohmic diffusion and Hall drift after a relatively long timescale \citep[e.g.,][]{1999A&A...345..847G, 2013ApJ...775..124F}. Given a re-diffusion timescale of $\gtrsim10^3$ years for $M_\mathrm{acc}>10^{-4}M_\odot$, the re-diffusion process of magnetic field is not considered in this paper.

\begin{figure*}[t]
\includegraphics[angle=0,width=0.5\textwidth]{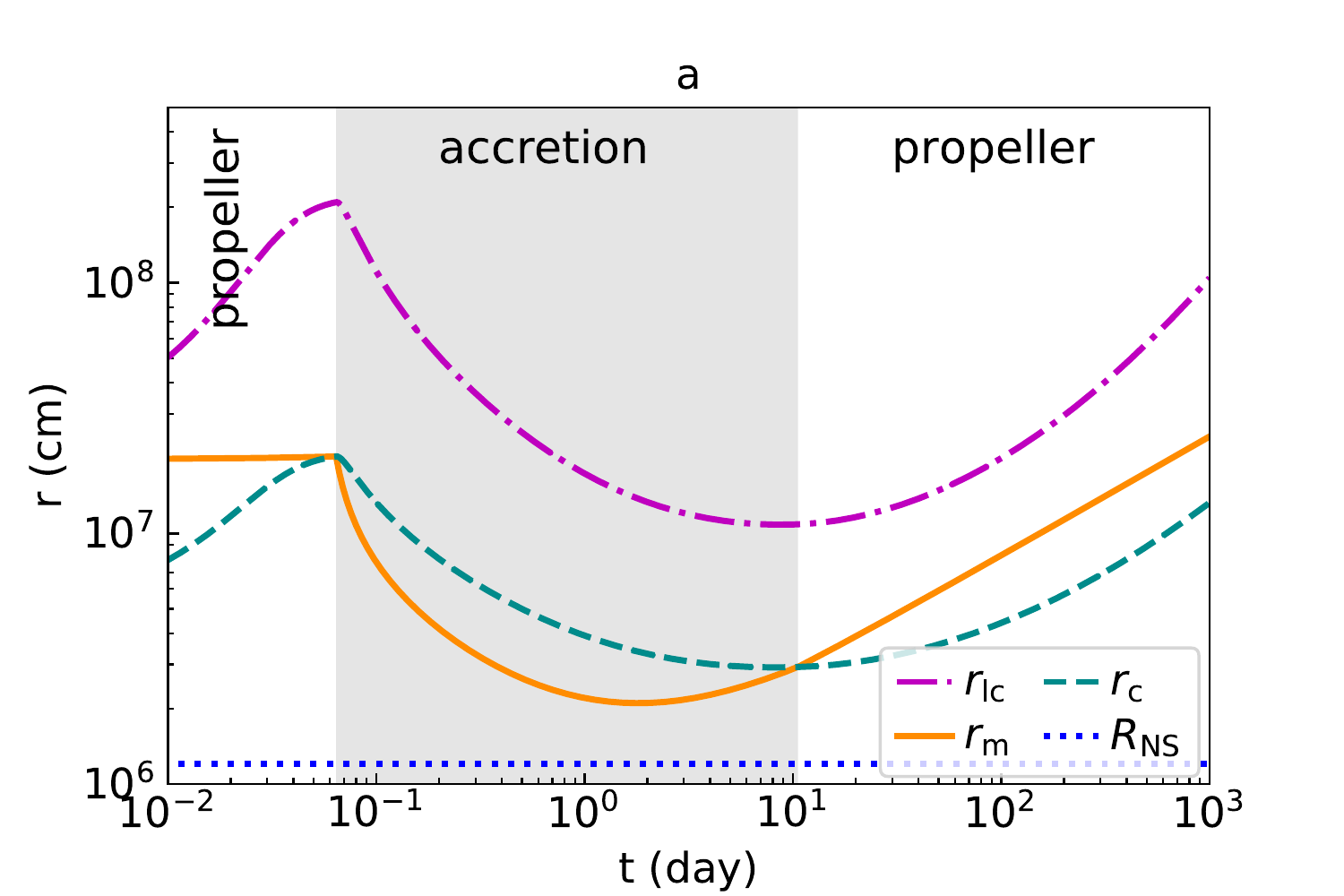}\includegraphics[angle=0,width=0.5\textwidth]{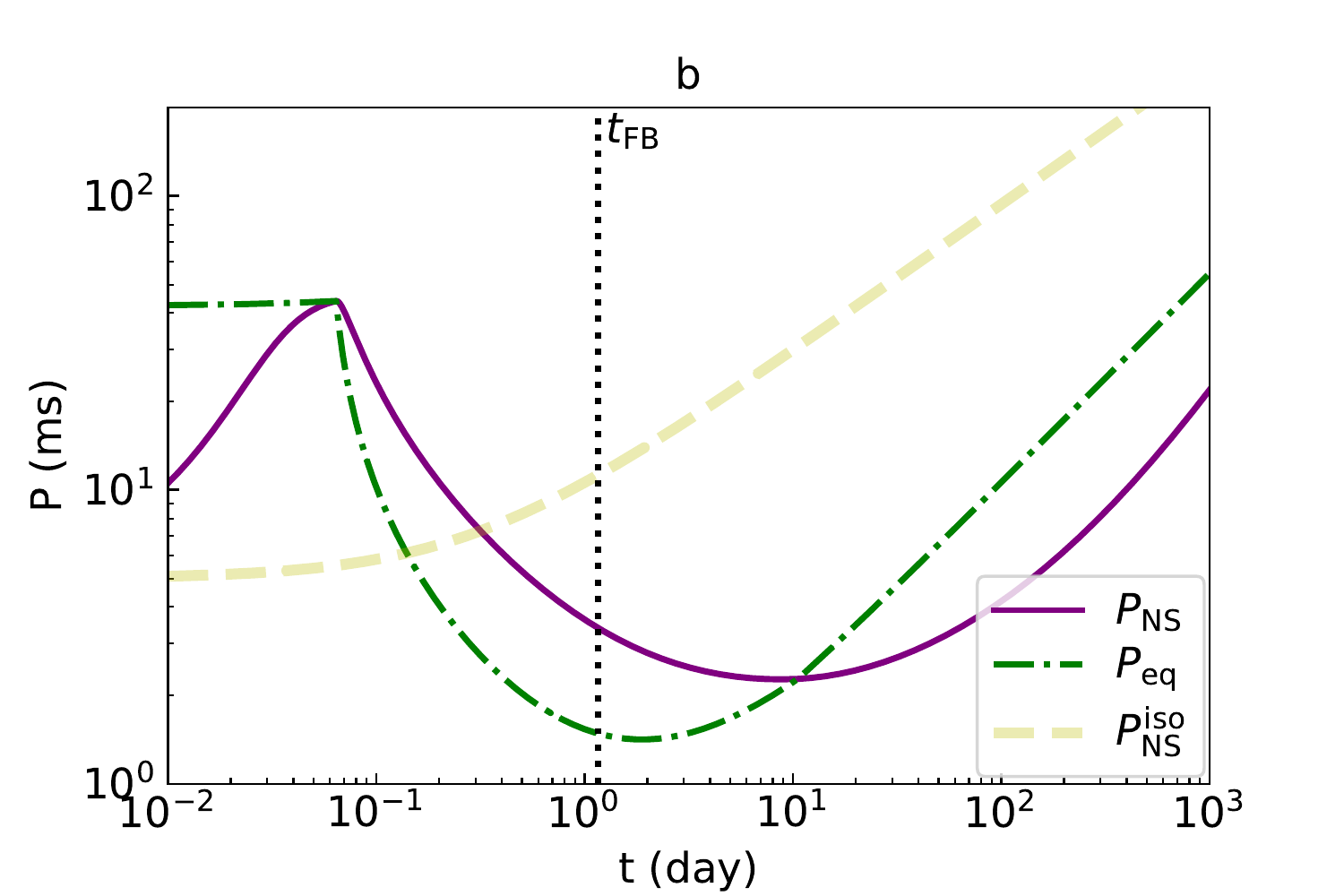}
\includegraphics[angle=0,width=0.5\textwidth]{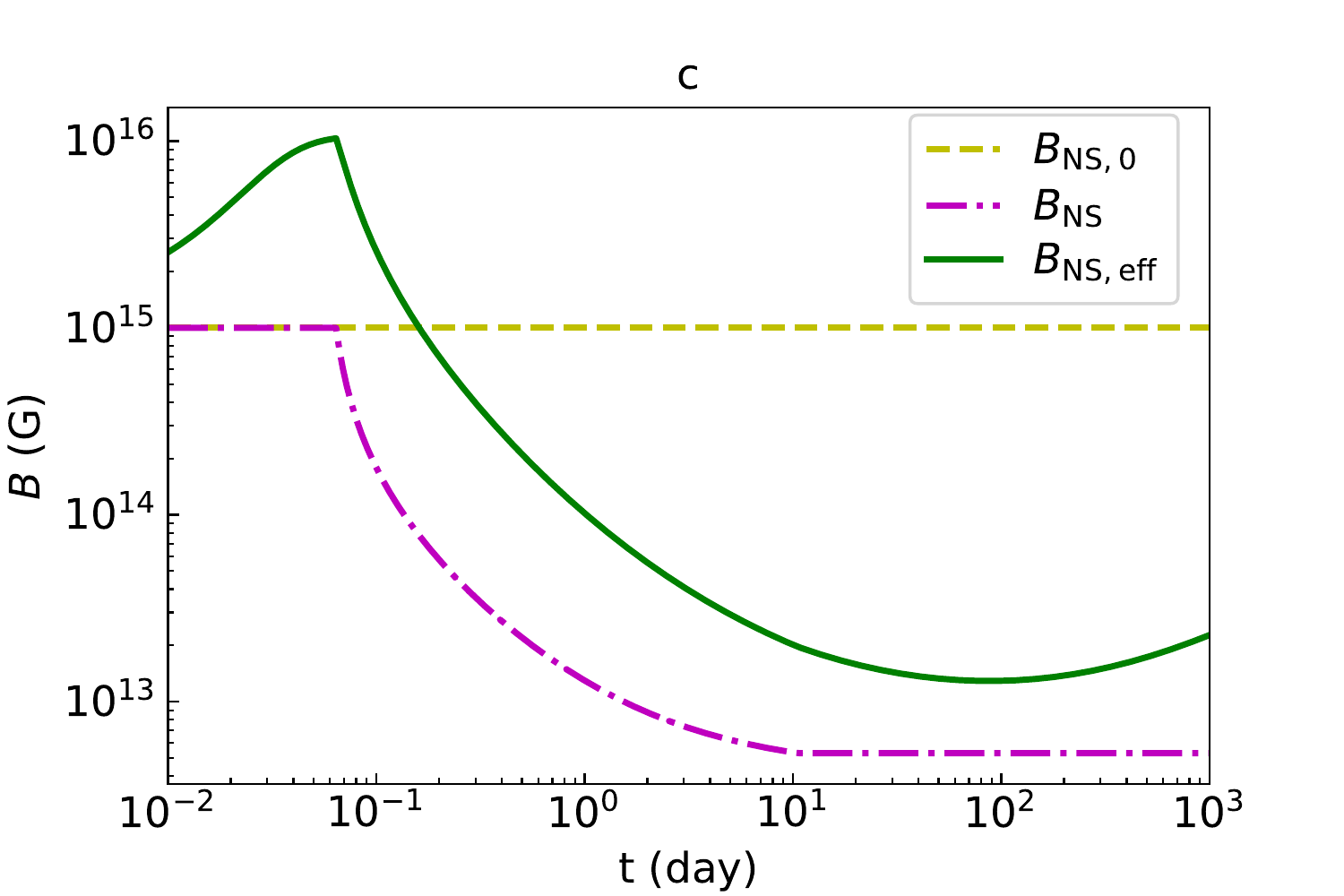}\includegraphics[angle=0,width=0.5\textwidth]{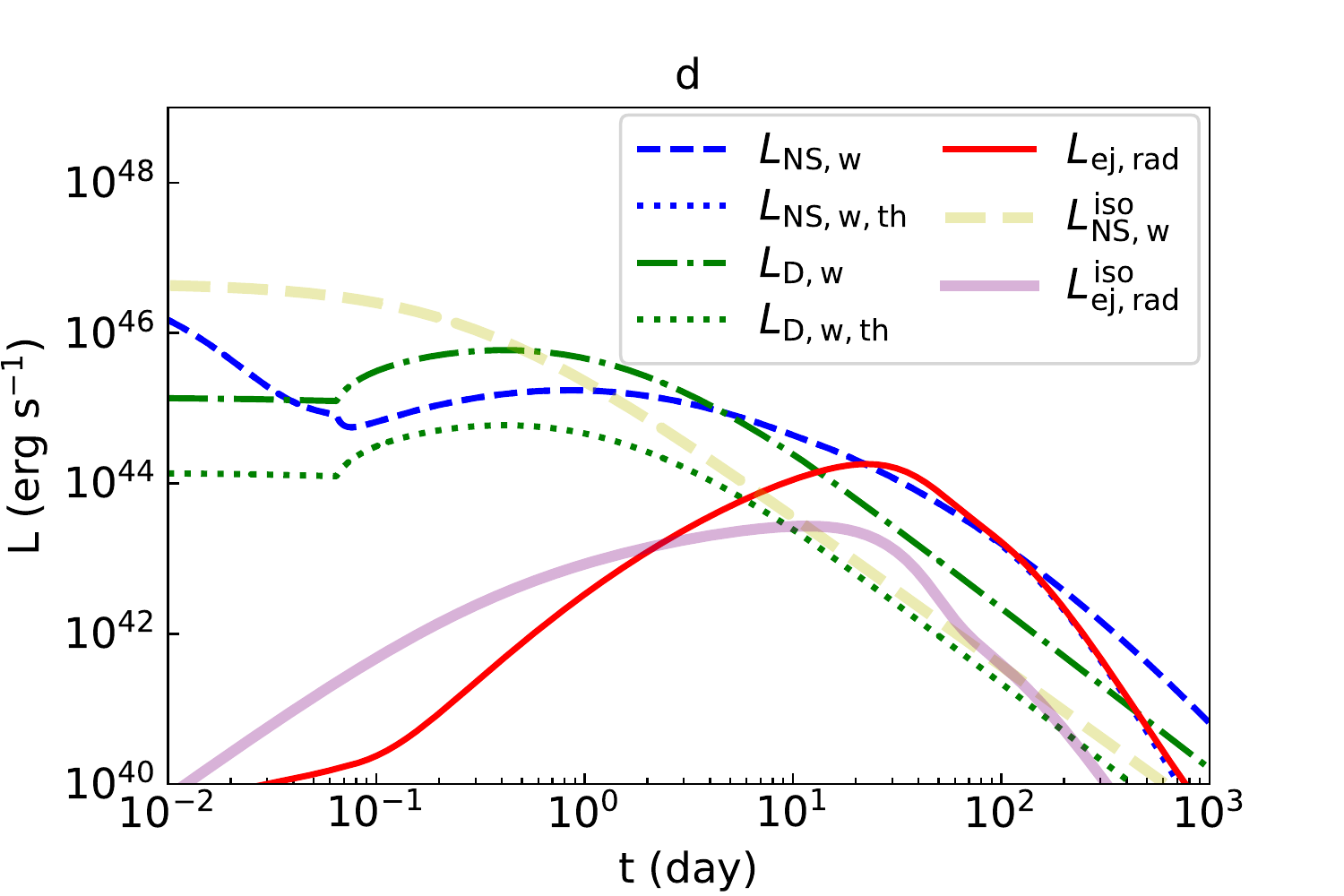}
\caption{The evolution of the magnetar-disk system for case A. (a) Characteristic radii: $r_\mathrm{lc}$ (dashed-dotted), $r_\mathrm{m}$ (solid), $r_\mathrm{c}$ (dashed) and $R_\mathrm{NS}$ (dotted). The accretion phase is shown by gray-shaded region. (b) The spin period $P_\mathrm{NS}$ (solid) and the accretion-induced equilibrium period $P_\mathrm{eq}$ (Equation \ref{eq: Peq}; dashed-dotted). The vertical dotted line indicates the fallback timescale. (c) Magnetic field: $B_\mathrm{NS,0}$ (dashed), $B_\mathrm{NS}$ (dashed-dotted) and $B_\mathrm{NS,eff}$ (solid). (d) The magnetar wind luminosity ($L_\mathrm{NS,w}$; blue dashed) and the fraction thermalized by the SN ejecta ($L_\mathrm{NS,w,th}$; blue dotted), the disk outflow luminosity ($L_\mathrm{D,w}$; green dashed-dotted) and the thermalized outflow luminosity ($L_\mathrm{D,w,th}$; green dotted), and the SN luminosity powered by this system ($L_\mathrm{ej,rad}$; red solid). For comparison, we show in panel (d) the magnetar wind luminosity ($L_\mathrm{NS,w}^\mathrm{iso}$, Equation \ref{eq: L_dip_iso}; yellow dashed) and the SN luminosity ($L_\mathrm{ej,rad}^\mathrm{iso}$, Equation \ref{eq: L_rad_iso}; purple solid) derived in the isolated magnetar-powered scenario with the same initial properties as case A, whose spin evolution ($P_\mathrm{NS}^\mathrm{iso}$, Equation \ref{eq: P_iso}; yellow dashed) is displayed in panel (b).}
\label{fig: 1_spin}
\end{figure*}

\subsection{SN luminosity powered by magnetar-disk system}
\label{Subsec: model_SN}

Magnetic dipole radiation can drive a magnetar wind with a luminosity of
\begin{equation}
L_\mathrm{NS,w}=\Omega_\mathrm{NS}N_\mathrm{dip}.
\label{eq: L_dip}
\end{equation}

The kinetic luminosity of large-scale outflow from the radiatively ineffective disk can be estimated by (see Appendix \ref{sec_app: disk} for detailed derivations)
\begin{equation}
L_\mathrm{D,w}\approx 0.001\eta\dot{M}_\mathrm{fb}c^2\left(\frac{M_\mathrm{NS}}{1.4M_\odot}\right)\left(\frac{r_\mathrm{m}}{10^7 \mathrm{cm}}\right)^{-1},
\label{eq: L_D}
\end{equation}

Outflow could be also generated from the inner disk during the propeller regime. However, as \citet{2021ApJ...907...87L} pointed out, the kinetic energy of the propeller outflow could be reduced because of (1) low acceleration efficiency, (2) internal dissipation inside the outflow, and (3) interaction between the outflows and the in-falling matter from the outer disk. Thus, the propeller outflow is not considered here. 

In the accretion regime, the highly super-Eddington accretion column above the magnetar should be radiatively inefficient and cool via neutrino emission \citep[e.g.,][]{2011ApJ...736..108P, 2018MNRAS.476.2867M}. Hence, the accretion luminosity is not expected to significantly affect the SN luminosity.

Assuming that an ejecta with mass $M_\mathrm{ej}$ and velocity $v_\mathrm{ej}$ is generated in SN explosion, the magnetar wind luminosity thermalized by the SN ejecta can be given by $L_\mathrm{NS,w,th}=(1-e^{-At'^{2}})L_\mathrm{NS,w}$, where $A=3\kappa_\mathrm{m} M_\mathrm{ej}/(4\pi v_\mathrm{ej}^2)$ is related to photon trapping \citep{2015ApJ...799..107W}, and $\kappa_\mathrm{m}$ is the opacity of SN ejecta to gamma-ray photons from magnetar wind. As for the mass outflow from the disk, a fraction of its kinetic energy can be used to heat the SN ejecta during the interaction process, i.e. $L_\mathrm{D,w,th}=\epsilon L_\mathrm{D,w}$, where $\epsilon$ is the thermalized efficiency. Then we use the semi-analytical solution for the bolometric light curve of SN ejecta in a homologous expansion derived by \citet{1982ApJ...253..785A} to calculate the SN luminosity powered by such a magnetar-disk system
\begin{equation}
L_\mathrm{ej,rad}(t)=2e^{-(t/t_\mathrm{diff})^2}\int^{t}_{0}(L_\mathrm{NS,w,th}+L_\mathrm{D,w,th})e^{(t'/t_\mathrm{diff})^2}\frac{t'\mathrm{d}t'}{t_\mathrm{diff}^2},
\label{eq: L_rad}
\end{equation}
where $t_\mathrm{diff}=(2\kappa M_\mathrm{ej}/(13.8 c v_\mathrm{ej}))^{1/2}$ is the diffusion time with $\kappa$ being the gray opacity of the SN ejecta. $\kappa$ can be constrained to $\sim0.01-0.2$ cm$^2$ g$^{-1}$ \citep{2013ApJ...770..128I}, while $\kappa_\mathrm{m}$ is usually assumed to be $0.01-100$ cm$^2$ g$^{-1}$ \citep[e.g.,][]{2017ApJ...850...55N}.

\section{Results}
\label{Sec: results}
Using the model described in Section \ref{Sec: model}, we further examine the effect of initial properties of the magnetar-disk system on the luminosity evolution of SNe within $1000$ days since explosion.

In case A, we assume (1) an SN ejecta with mass $M_\mathrm{ej}=5M_\odot$ and velocity $v_\mathrm{ej}=10^9$ cm s$^{-1}$, (2) a magnetar born with initial mass $M_\mathrm{NS,b,0}=1.4M_\odot$, spin period $P_\mathrm{NS,0}=5$ms, and magnetic field $B_\mathrm{NS,0}=10^{15}$ G, and (3) a fallback accretion disk with a total mass $M_\mathrm{fb}=0.5M_\odot$, fallback timescale $t_\mathrm{fb}=10^5$ s. The thermalized efficiency of disk outflow is $\epsilon=0.1$, and both opacities ($\kappa$ and $\kappa_\mathrm{m}$) are adopted as $0.1$ cm$^2$ g$^{-1}$.

As seen in Figure \ref{fig: 1_spin}, this system experiences three evolution stages within 1000 days, i.e. propeller ($t<0.06$ days), accretion ($t=0.06-11$ days), and propeller ($t>11$ days). During the first propeller period, the magnetar contributes its angular momentum to the disk, resulting in an increase in $r_\mathrm{c}$. After $t\approx0.06$ days, this system enter into the accretion regime ($r_\mathrm{c}>r_\mathrm{m}$). As the disk matter is accreted onto the surface of the magnetar, the magnetar spins up, grows in mass and declines in magnetic field strength. Since $r_\mathrm{m}\propto M_\mathrm{NS}^{-1/7}B_\mathrm{NS}^{8/7}$, the inner radius of the disk starts to shrink rapidly. As $t>t_\mathrm{fb}=10^5$ s (i.e. 1.2 days), the disk undergoes a substantial decreases in mass inflowing rate (i.e. $\dot{M}_\mathrm{fb}\propto t^{-5/3}$). Consequently, the ram pressure of the inflows decreases significantly, and the magnetic pressure of the magnetar pushes the disk outwards. After $t\approx11$ days, $r_\mathrm{m}$ exceeds the co-rotation radius and the propeller mechanism starts to work again. During this period, the magnetar spins down, and its magnetic field ceases to decay since the magnetar mass remains constant. Throughout the evolution of $t=0-1000$ days, the effective magnetic field is always enhanced to be above $B_\mathrm{NS}$ by the fallback accretion disk because $r_\mathrm{m}< r_\mathrm{lc}$. Nevertheless, $B_\mathrm{NS,eff}$ declines below the initial magnetic field after $t\sim0.1$ days due to accretion-induced $B_\mathrm{NS}$ decay. Although low $B_\mathrm{NS,eff}$ can weaken the magnetar wind, the energy transfer from the disk during the accretion regime results in spin-up of the magnetar and then significantly boost the magnetar wind. Before $t\sim100$ days, magnetar wind can be completely thermalized by the SN ejecta. However, when the SN ejecta becomes transparent due to expansion, only a fraction of wind luminosity can contribute to the SN luminosity. Since the kinetic luminosity of disk outflow is lower than the magnetar wind luminosity during $t>10$ days, the thermalized luminosity $L_\mathrm{D,w,th}$ is well below $L_\mathrm{NS,w,th}$ given the thermalized efficiency $\epsilon=0.1$. Powered by this magnetar-disk system, the SN exhibits a peak luminosity ($L_\mathrm{ej,rad,p}=2\times10^{44}$ erg s$^{-1}$) similar to those of SLSNe I. It is much more luminous than that powered by an isolated magnetar with the same initial mass, spin period and magnetic field.

\begin{figure*}[t]
\includegraphics[angle=0,width=0.5\textwidth]{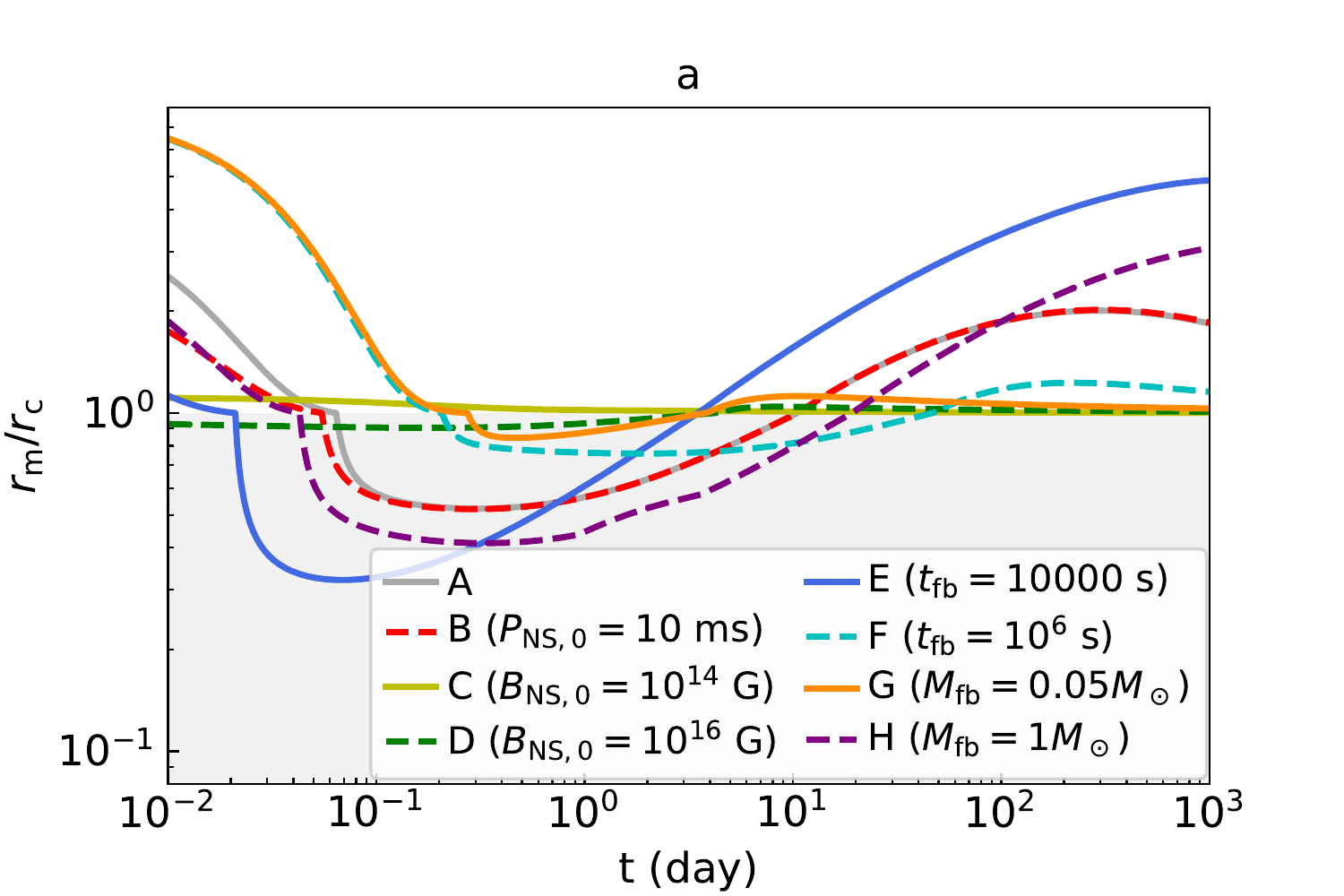}\includegraphics[angle=0,width=0.5\textwidth]{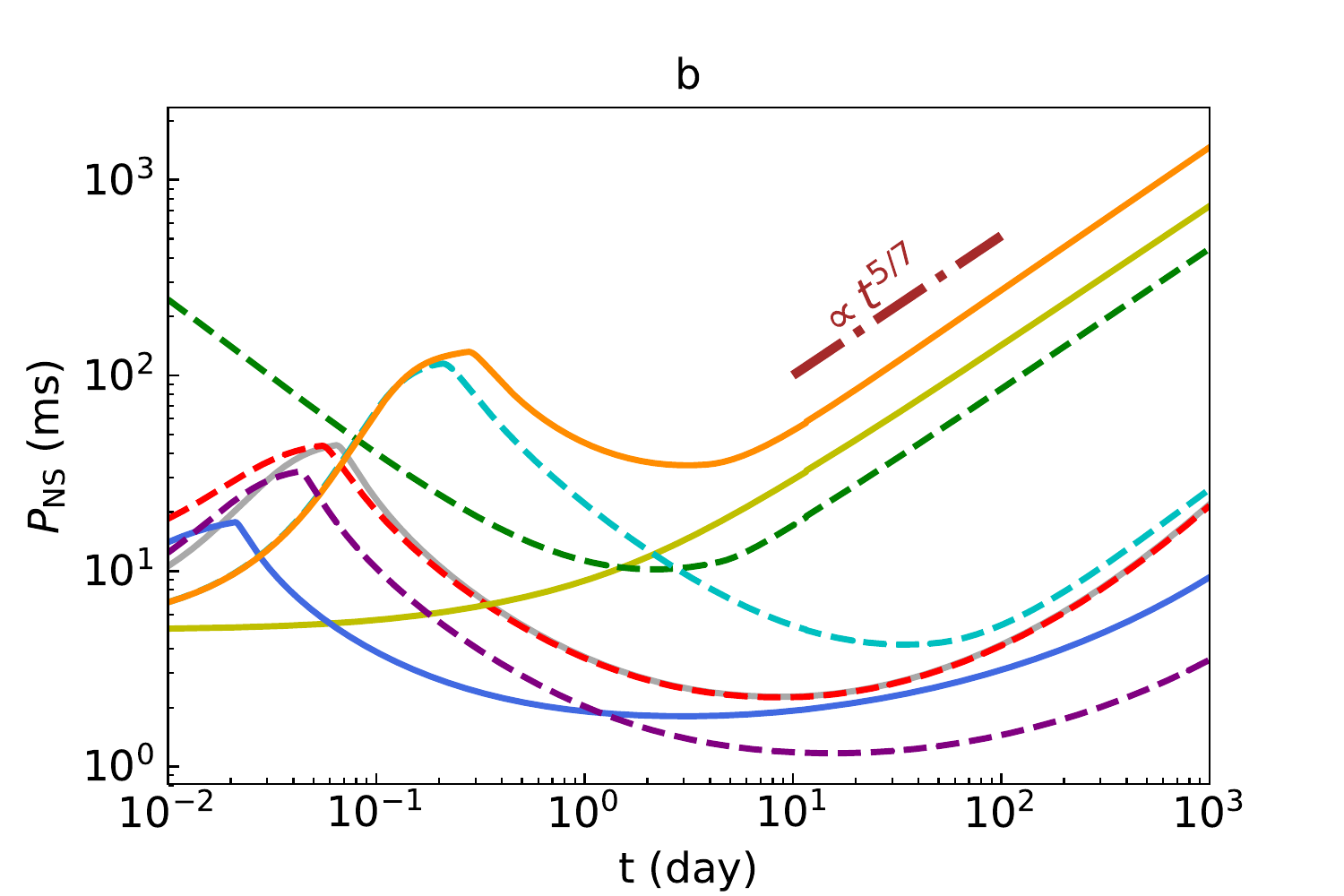}
\includegraphics[angle=0,width=0.5\textwidth]{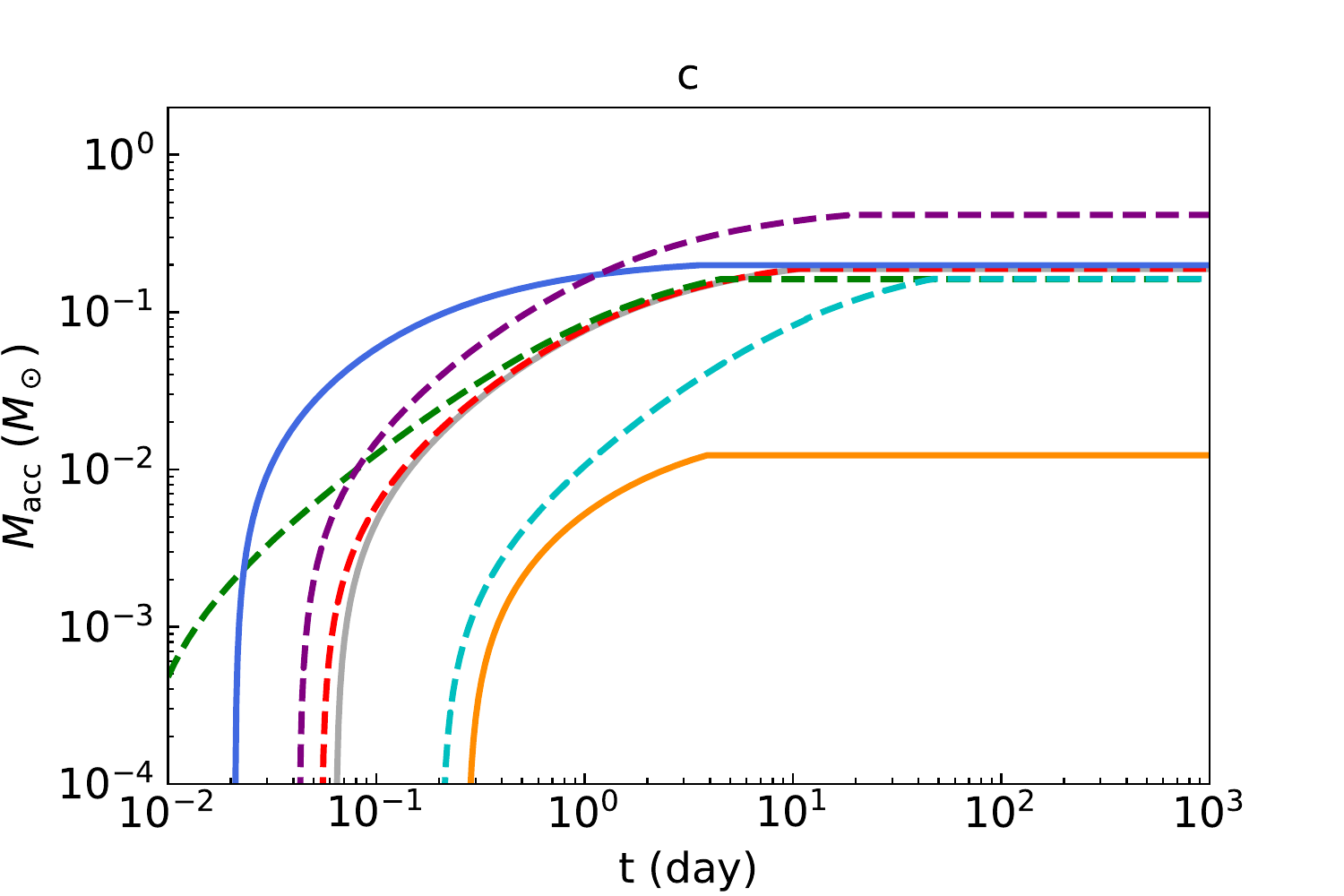}\includegraphics[angle=0,width=0.5\textwidth]{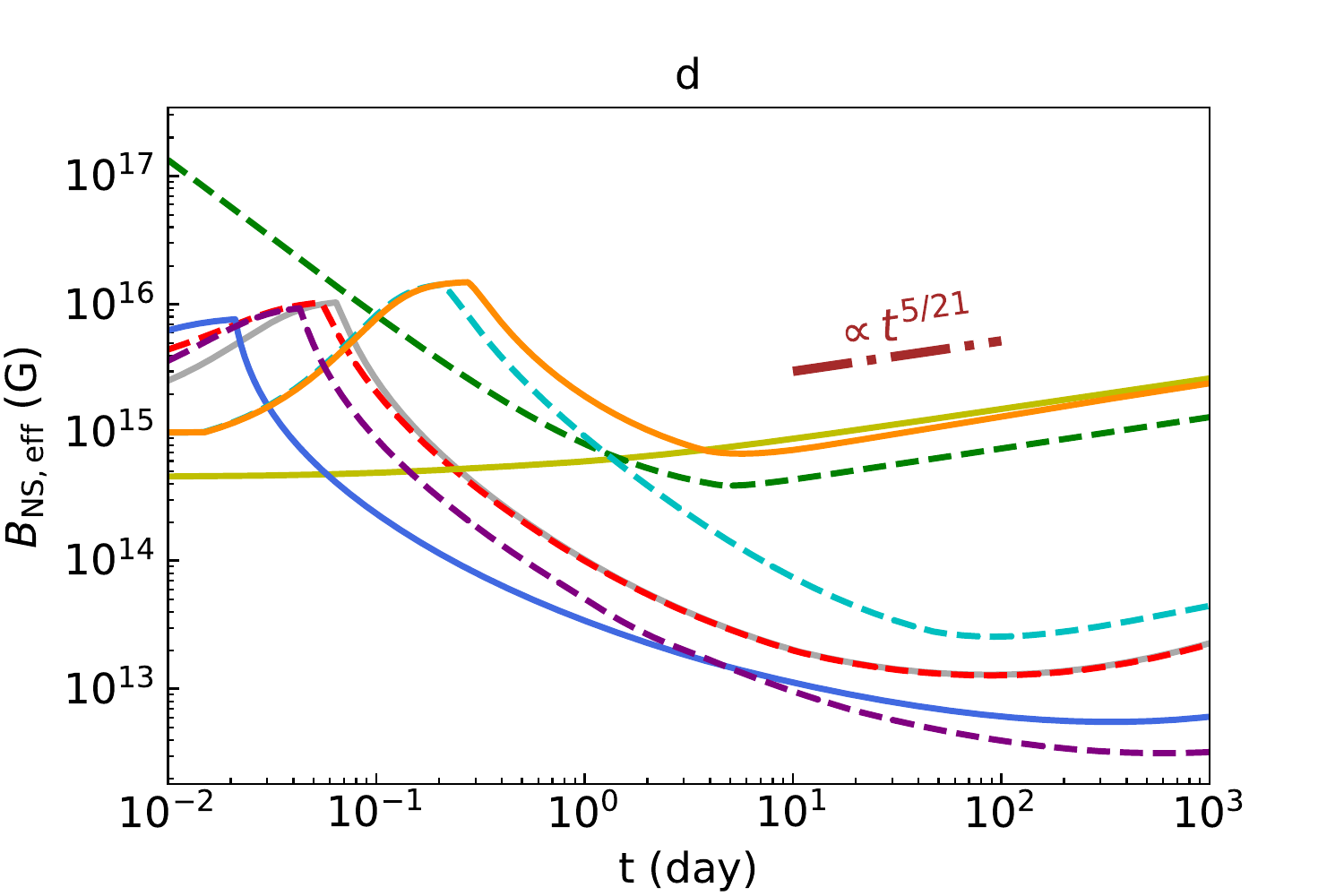}
\includegraphics[angle=0,width=0.5\textwidth]{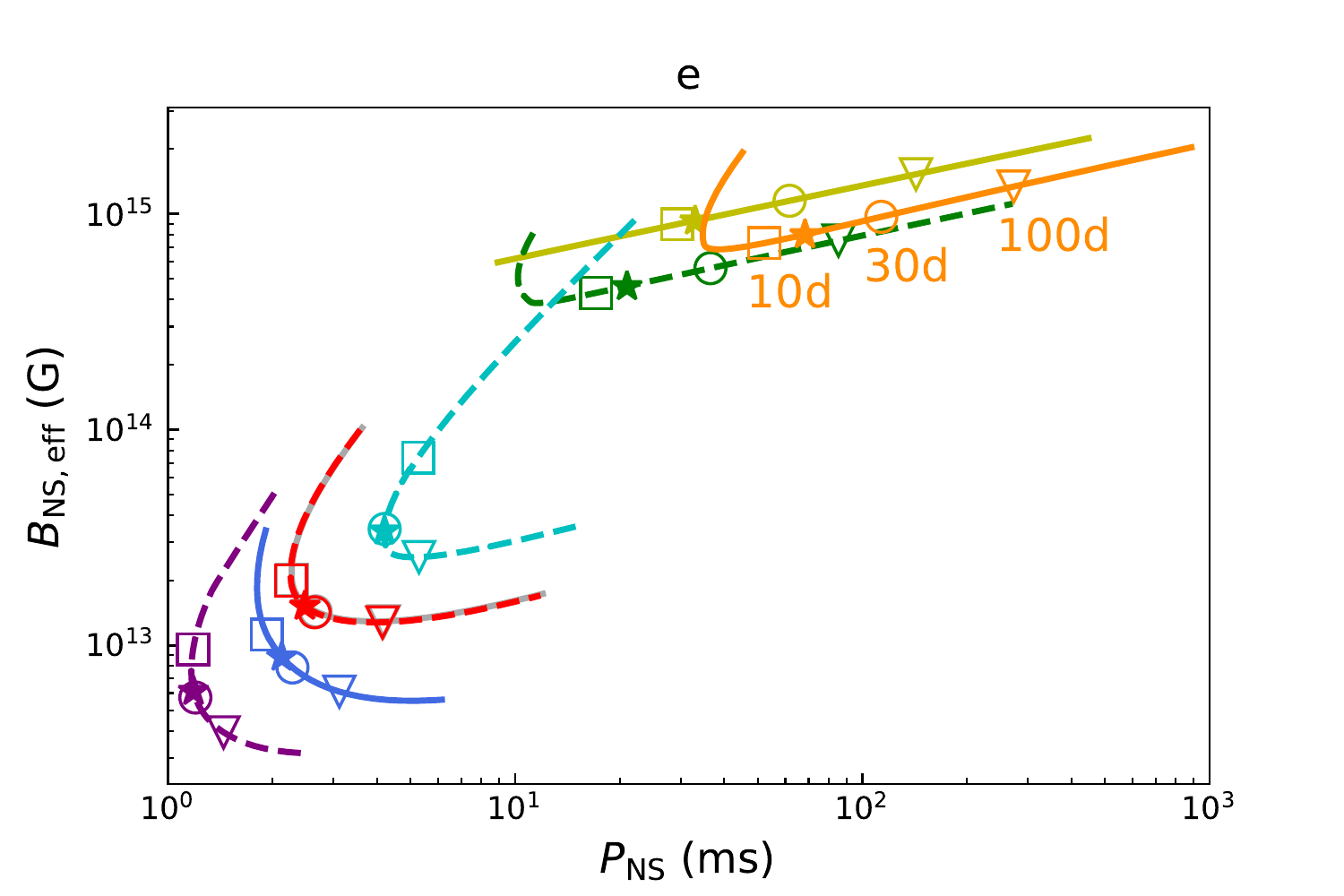}\includegraphics[angle=0,width=0.5\textwidth]{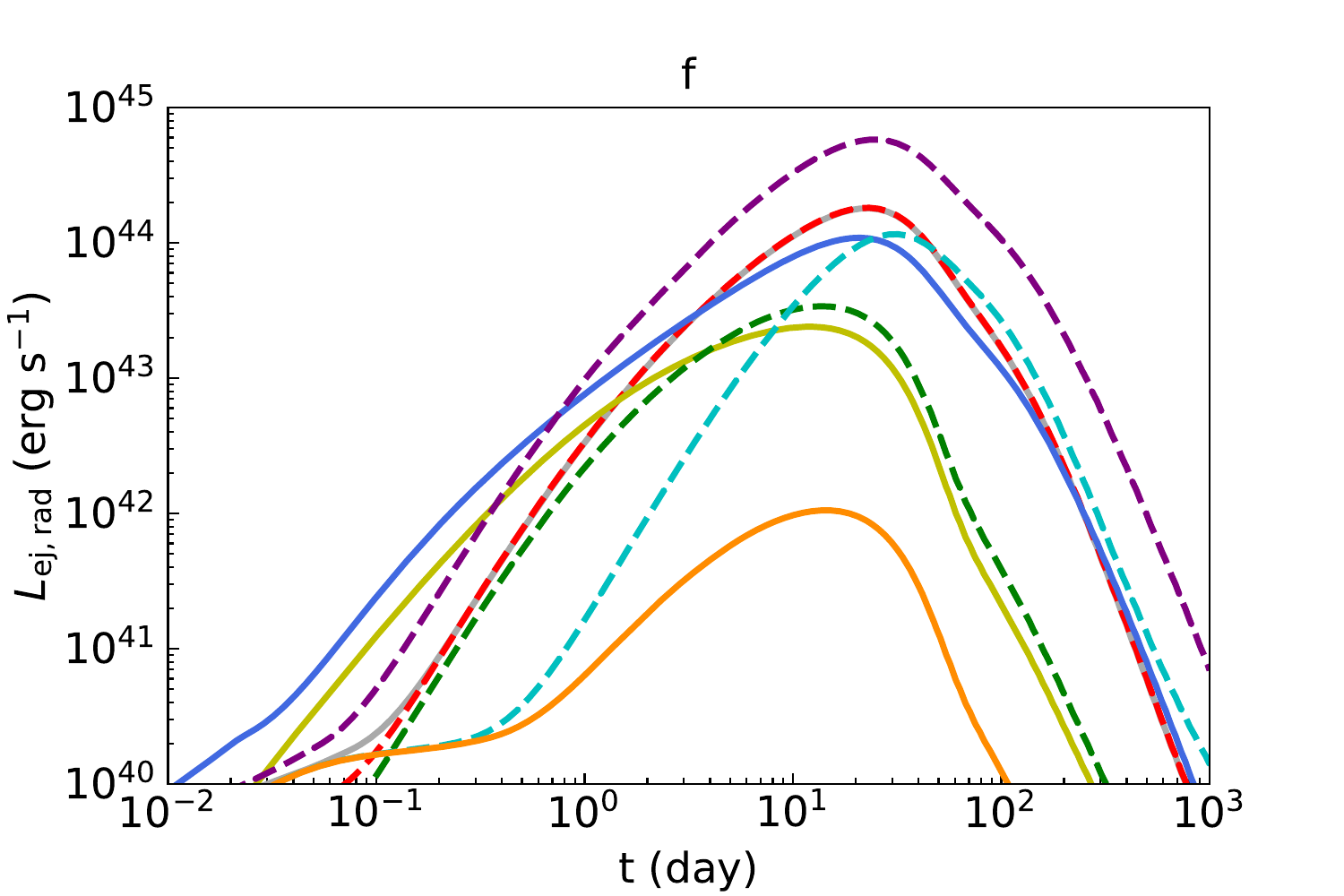}
\caption{Comparison of the evolution of the magnetar-disk systems with different initial properties. In cases B--H, only an initial parameter is assumed to be different from that adopted in case A and the rest parameters remain the same. We show the radii ratio $r_\mathrm{m}/r_\mathrm{c}$ (a), the spin period of the magnetar $P_\mathrm{NS}$ (b), the accretion mass $M_\mathrm{acc}$ (c), the effective magnetic field $B_\mathrm{NS,eff}$ (d), $B_\mathrm{NS,eff}-P_\mathrm{NS}$ distribution during $t=1-500$ days (e) and the SN light curves powered by such systems (f). In panel (e), we mark the $B_\mathrm{NS,eff}-P_\mathrm{NS}$ distribution at $t=10$ (squares), 30 (circles) and 100 days (triangles) as well as the peak time (stars) of SNe. For a system that evolves at an equilibrium state during $t>t_\mathrm{fb}$, the spin period of the magnetar evolves as $P_\mathrm{NS}|_\mathrm{eq}\propto t^{5/7}$, and the effective magnetic field is $B_\mathrm{NS,eff}|_\mathrm{eq}\propto t^{5/21}$ (see the dotted-dashed lines in panels b and d). See text for detailed discussions.}
\label{fig: Comp.}
\end{figure*}

\begin{figure}[thb]
\includegraphics[angle=0,width=0.5\textwidth]{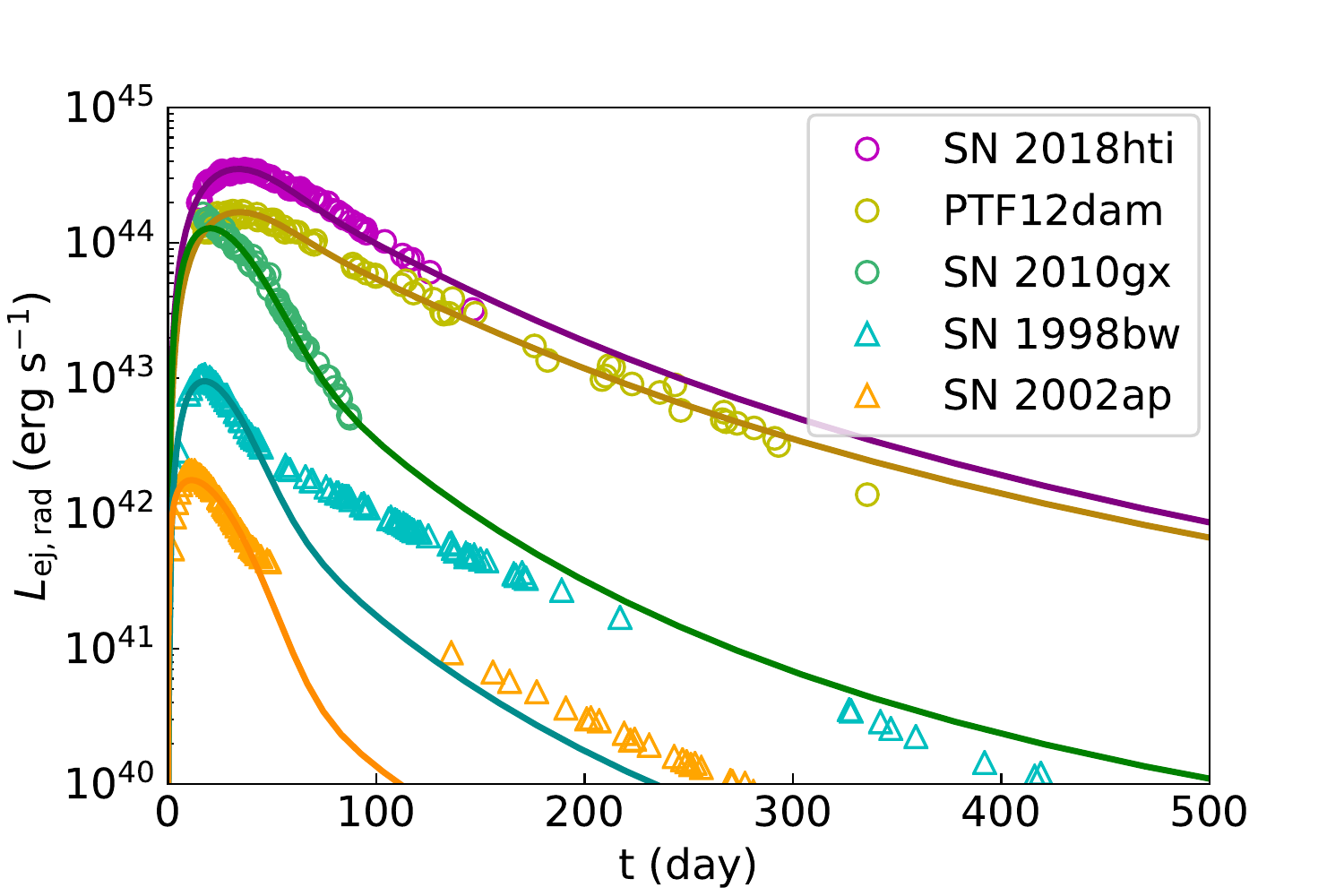}
\caption{The bolometric light curves of some representative SLSNe I (circles) and SNe Ic-BL (triangles) with respect to artificially assumed explosion epoch. The bolometric luminosities of SLSNe I are derived by fitting the absorbed blackbody curve \citep{2017ApJ...850...55N} to the multi-band data \citep{2013Natur.502..346N, 2014Ap&SS.354...89B, 2017ApJ...835...64G, 2018ApJ...860..100D, 2020MNRAS.497..318L}, while the bolometric light curves of SN 1998bw and SN 2002ap are taken from \citet{2001ApJ...555..900P} and \citet{2006ApJ...644..400T}, respectively. The solid lines represent the fitting results from our model.}
\label{fig: Comp_Lrad}
\end{figure}

In Figure \ref{fig: Comp.}, we take case A as the basic scenario and vary only one initial parameter in our simulations for case B--H to study the effect of initial parameters on the evolution of magnetar-disk system and on the luminosity of the SNe. When we assume $P_\mathrm{NS,0}=1-10$ ms, the evolution of the system (see case B as an example with $P_\mathrm{NS,0}=10$ ms) after $t=0.1$ days is similar to that in case A, suggesting that the initial spin period of the magnetar might not have significant influence on the late evolution of the system. 

The evolution of $r_\mathrm{m}/r_\mathrm{c}$ is shown in Figure \ref{fig: Comp.}a. The magnetar with $B_\mathrm{NS,0}=10^{14}$ G (case C) is always in the propeller regime ($r_\mathrm{m}/r_\mathrm{c}>1$) while the other systems considered here experience propeller--accretion--propeller stages. Both the accretion and the magnetic dipole torques influence the spin evolution of the magnetar, but the former plays a dominant role during most of the evolution time of our interest in above cases. Thus, the central magnetar usually spins up in the accretion phase but slows down in the propeller phase (Figure \ref{fig: Comp.}b). Compared to case A and F ($t_\mathrm{fb}=10^6$ s), the accretion phase starts and ends earlier when the system has stronger initial magnetic field $B_\mathrm{NS,0}=10^{16}$ G (case D) or shorter fallback timescale $t_\mathrm{fb}=10^4$ s (case E). Nevertheless, total accretion masses in these four cases are comparable (Figure \ref{fig: Comp.}c). When the fallback mass varies between $0.05-1M_\odot$, we find that the accretion phase could start earlier and last for a longer time for the system with a larger fallback mass. Moreover, since the fallback timescale is assumed to be the same in cases A, G ($M_\mathrm{fb}=0.05M_\odot$) and H ($M_\mathrm{fb}=1M_\odot$), larger fallback mass corresponds to higher mass inflowing rate, which results in a larger mass accreted onto the surface of the magnetar (Figure \ref{fig: Comp.}c). According to Equation (\ref{eq: B_NS}), only the magnetar in case C that keeps expelling matter from the disk possesses a constant field throughout the evolution; while in other cases, the magnetic field of magnetars decays significantly due to mass accretion and then remains invariable after the accretion regime ends. The effective magnetic field strength ($B_\mathrm{NS,eff}$) can be enhanced when the disk penetrates the light cylinder of the magnetar ($r_\mathrm{m}<r_\mathrm{lc}$); but it might become overall weak during the accretion stage if the accretion-induced magnetic field decays significantly (Figure \ref{fig: Comp.}d).

It is worth noting that the magnetar wind power (see Equation \ref{eq: L_dip}) is determined by the spin period and the effective magnetic field ($B_\mathrm{NS,eff}$), instead of the magnetic field ($B_\mathrm{NS}$). In Figure \ref{fig: Comp.}e, we show the $B_\mathrm{NS,eff}-P_\mathrm{NS}$ distribution during 1-500 days after explosion. In cases C, D and G, the magnetars rotate with $P_\mathrm{NS,p}=10-100$ ms around the epoch of maximum light ($t_\mathrm{p}$), and their $B_\mathrm{NS,eff,p}$ is enhanced to $\sim10^{14}-10^{15}$ G by the disk. We note that, during $t=10-1000$ days, these three systems are all in the propeller regime and evolve at a near-equilibrium state with $P_\mathrm{NS}|_\mathrm{eq}\propto t^{5/7}$ and $B_\mathrm{NS,eff}|_\mathrm{eq}\propto t^{5/21}$ (see dotted-dashed lines in Figure \ref{fig: Comp.}b and \ref{fig: Comp.}d). In the other five cases, however, the magnetar engines are characterized by lower effective field ($B_\mathrm{NS,eff,p}=10^{12}-10^{14}$ G) and faster spin ($P_\mathrm{NS,p}=1-10$ ms) around $t_\mathrm{p}$. Therefore, there seems to be a positive correlation between $B_\mathrm{NS,eff,p}$ and $P_\mathrm{NS,p}$ at peak, which is reminiscent of the positive correlation between $B_\mathrm{NS}$ and $P_\mathrm{NS,0}$ inferred from isolated magnetar model for SLSNe I and SNe Ic-BL \citep{2020ApJ...903L..24L}.

In Figure \ref{fig: Comp.}f, we display the bolometric light curves of the SNe powered by these magnetar-disk systems. The thermalized luminosity of disk outflow is always lower than that of the magnetar wind in these cases. The peak luminosities of these SNe can vary from $10^{42}$ erg s$^{-1}$ to $10^{45}$ erg s$^{-1}$, which cover the values observed in SNe Ic/Ic-BL ($41.5\lesssim\log L_\mathrm{p}\lesssim43.5$; e.g., \citealp{2016MNRAS.458.2973P}) and SLSNe I ($\log L_\mathrm{p}>43.5$; e.g., \citealp{2019NatAs...3..697I}). In cases A--H, SNe with $L_\mathrm{p}=10^{42}-4\times10^{43}$ erg s$^{-1}$ reach the peak luminosity at $t_\mathrm{p}=11-15$ days since explosion, while a much higher peak luminosity (i.e. $L_\mathrm{p}>10^{44}$ erg s$^{-1}$) can be attained in a light curve with a longer rise time (i.e. 20--32 days). Thus, a positive correlation likely exists between the peak luminosity and the rise time, which is in agreement with the observation tendency that SLSNe I have broader and brighter light curves than SNe Ic-BL. 

As seen in Figure \ref{fig: Comp_Lrad}, our model with the parameters listed in Table \ref{Tab: fit} can reproduce the observed light curves of some representative fast- and slow-evolving SLSNe I (i.e., PTF12dam, SN 2010gx and SN 2018hti). The early-time light curves of SNe Ic-BL (i.e., SN 1998bw and SN 2002ap) can be also explained in terms of the magnetar-disk interaction scenario. However, we notice that the late-time luminosities of SN 1998bw and SN 2002ap appear to be much higher than the theoretical light curves, which is possibly due to the contribution of $^{56}$Ni powering \citep{2017ApJ...851...54W, 2017ApJ...837..128W}. Thus, magnetar wind regulated by the magnetar-disk interaction can serve as a primary power source for both SLSNe I and SNe Ic-BL.

\begin{table}[htb]
 \centering
  \caption{Model fitting parameters for some well-observed SNe Ic-BL and SLSNe I. We set $M_\mathrm{NS,b,0}=1.4M_\odot$, $P_\mathrm{NS,0}=5$ ms, $M_\mathrm{ej}=5M_\odot$, $v_\mathrm{ej}=10000$ km s$^{-1}$, $\epsilon=0.1$, and $\kappa_\mathrm{m}=0.1$ cm$^2$ g$^{-1}$. \label{Tab: fit}}
 \setlength{\tabcolsep}{1mm}{
  \begin{tabular}{cccccc}
  \hline
  SN     &   type&    $B_\mathrm{NS,0}$ (G) & $M_\mathrm{fb}$ ($M_\odot$)  & $t_\mathrm{fb}$ (s) & $\kappa$ (cm$^2$ g$^{-1}$)\\
  \hline
  SN 1998bw  & Ic-BL       &     $8.5\times10^{14}$               & 0.12   & $10^5$   &  0.1 \\
  SN 2002ap   & Ic-BL     &     $3\times10^{14}$    & 0.2 &  $5\times10^4$  &  0.13 \\
  PTF12dam    &SLSN I     &     $5\times10^{14}$    & 0.67   &   $10^5$    &  0.2 \\
  SN 2010gx     &SLSN I    &     $2.8\times10^{15}$    & 0.6 &  $10^5$  &  0.15 \\
  SN 2018hti      &SLSN I    &     $9\times10^{14}$    & 0.92     &  $10^5$  &  0.2 \\
   \hline
\end{tabular}}
\\{Note: We present one but not unique set of parameters for modeling the light curve of each SN.}
\end{table}

Since our model may reproduce the major observational characteristics of the near-maximum-light bolometric light curves of SLSNe I or SNe Ic-BL, these two subclasses of SNe could have a similar origin, which is also implied by the similarity between their late-time spectra \citep[e.g.,][]{2010ApJ...724L..16P, 2017ApJ...845...85L, 2019ApJ...872...90B, 2019ApJ...871..102N, 2020MNRAS.497..318L}. As for the observed differences in their early-time spectra, magnetar wind that remains powerful for tens of days since explosion could help produce the prominent O\,\textsc{ii} absorption features seen in SLSNe I instead of SNe Ic-BL, via non-thermal excitation or by heating the SN ejecta to a high temperature \citep{2011Natur.474..487Q, 2018ApJ...855....2Q, 2016MNRAS.458.3455M}.

Finally, we give a brief discussion over the possible connection between SLSNe I and long gamma-ray bursts (LGRBs). Rapidly-rotating magnetars have been proposed as one of the promising central engines for gamma-ray bursts \citep[e.g.,][]{1992Natur.357..472U, 1998A&A...333L..87D, 2001ApJ...552L..35Z}. As shown in Figure 1 of \citet{2020ApJ...903L..24L}, strong magnetic field ($>10^{14}$ G) might play a crucial role in driving a magnetar wind responsible for the shallow decay of early-time afterglow of LGRBs, while relatively low magnetic field strength ($\lesssim$ a few $10^{14}$ G) is required for the isolated millisecond magnetars to power the broad and luminous light curves of SLSNe I that peak at tens of days after the SN explosions. Thus, most LGRBs are not expected to be associated with SLSNe I in the isolated magnetar-powered scenario. However, their association, if observed in the future, can be explained in the context of magnetar-disk system where the magnetic field of the nascent magnetar could decay significantly due to fallback accretion, since the magnetar-disk scenario is also favored for some LGRB afterglows \citep[e.g.,][]{2012ApJ...759...58D, 2021ApJ...907...87L}.

\section{Conclusion}
\label{Sec: conclusion}

In this paper, we study the effect of fallback accretion on the evolution of central magnetar and SN luminosity. On one hand, fallback accretion might accelerate the spin of the magnetar in the accretion regime, and then the SN ejecta is heated by stronger magnetar wind. On the other hand, the SN luminosity can be low, when the magnetar spins down substantially during the propeller regime. The main conclusions are outlined in the following.

Firstly, in the presence of a fallback accretion disk, the evolutions of the magnetar and the SN luminosity depend strongly on the magnetic field of the magnetar as well as the fallback mass and timescale for the disk, while the initial spin period of the magnetar plays a less significant role. 

Secondly, light curves of both SNe Ic-BL and SLSNe I can be reproduced in the magnetar-disk interaction scenario, suggesting that these two subclasses of SNe could have a similar origin. Compared to the magnetars in SNe Ic-BL, those that can power SLSNe I usually maintain faster rotation and relatively lower effective magnetic field around the light-curve peak time.

Finally, we revisit the possible link between LGRBs and SLSNe I in the context of magnetar-disk system. Fallback accretion could result in a significant decay in the magnetic field of a millisecond magnetar born with strong magnetic field that is required for LGRBs, which makes it possible for the magnetar to power an energetic SN similar to SLSNe I at tens of days after explosion.

 \section*{Acknowledgements}

The authors thank the anonymous referee for his/her suggestive comments that help improve the paper. This work is supported by the National Natural Science Foundation of China (NSFC grants 12033003, 11633002, 11761141001 and 11833003), the National Program on Key Research and Development Project (grant 2016YFA0400803 and 2017YFA0402600), the Scholar Program of Beijing Academy of Science and Technology (DZ:BS202002), and the National SKA Program of China (grant No. 2020SKA0120300). L.J.W. acknowledges support from the National Program on Key Research and Development Project of China (grant 2016YFA0400801).

{}

\appendix
\section{Outflow luminosity from disk}
\label{sec_app: disk}
In this paper, we assume the accretion rate of the disk as a power-law function of radius with a constant index $0<s<1$ \citep{2005ApJ...629..341K}, 
\begin{equation}
\dot{M}_\mathrm{D}(r)=\dot{M}_{\mathrm{fb}}\left(r/r_\mathrm{out}\right)^{s},
\label{eq: Mdot_r}
\end{equation}
where $r_\mathrm{out}$ is the outer radius of the disk. In this case, the accretion rate ratio $\eta=\dot{M}_\mathrm{D,in}/\dot{M}_{\mathrm{fb}}=\left(r_\mathrm{m}/r_\mathrm{out}\right)^{s}$. Given that the velocity of the large-scale outflow from the disk is likely to be comparable to the local escape velocity $v_\mathrm{es}=(2G M_\mathrm{NS}/r)^{1/2}=(r_\mathrm{S}/r)^{1/2}c$, the kinetic luminosity of the outflow can be estimated by \citep{2005ApJ...629..341K}
\begin{equation}
\begin{aligned}
L_\mathrm{D,w}&\approx \frac{1}{2}\zeta\int^{r_\mathrm{out}}_{r_\mathrm{m}} v_\mathrm{es}^2\mathrm{d}\dot{M}_\mathrm{D}(r)\\
&=\frac{\zeta s}{2(1-s)}\frac{r_\mathrm{S}}{r_\mathrm{m}}\left[\left(\frac{r_\mathrm{m}}{r_\mathrm{out}}\right)^{s}-\frac{r_\mathrm{m}}{r_\mathrm{out}}\right]\dot{M}_\mathrm{fb}c^2\\
&\lesssim\frac{\zeta s}{2(1-s)}\frac{r_\mathrm{S}}{r_\mathrm{m}}\eta\dot{M}_\mathrm{fb}c^2\\
&\lesssim0.002\eta\dot{M}_\mathrm{fb}c^2\left(\frac{s}{1-s}\right)\left(\frac{\zeta}{0.1}\right)\left(\frac{M_\mathrm{NS}}{1.4M_\odot}\right)\left(\frac{r_\mathrm{m}}{10^7 \mathrm{cm}}\right)^{-1}
\end{aligned}
\label{eq: L_D_r}
\end{equation}
where $r_\mathrm{S}=2G M_\mathrm{NS}/c^2$ is Schwarzschild radius, and $\zeta$ parameterizes the effect of outflow physics. In this paper, we adopt (see also Equation \ref{eq: L_D})
\begin{equation}
L_\mathrm{D,w}\approx 0.001\eta\dot{M}_\mathrm{fb}c^2\left(\frac{M_\mathrm{NS}}{1.4M_\odot}\right)\left(\frac{r_\mathrm{m}}{10^7 \mathrm{cm}}\right)^{-1},
\label{eq: L_D_r1}
\end{equation}
instead of $L_\mathrm{D,w}\approx 0.001\dot{M}_\mathrm{fb}c^2$ used in the fallback accretion-powered model \citep{2013ApJ...772...30D}. 

\section{Isolated magnetar-powered SNe}

For an isolated magnetar, the rotation energy is dissipated via the magnetic dipole radiation. The spin and the wind luminosity (i.e. magnetic dipole luminosity) of the magnetar can be written as
\begin{equation}
P_\mathrm{NS}^\mathrm{iso}=P_\mathrm{NS,0}(1+t/T_\mathrm{dip}^\mathrm{iso})^{1/2},
\label{eq: P_iso}
\end{equation}
\begin{equation}
L_\mathrm{NS,w}^\mathrm{iso}=\frac{B_\mathrm{NS,0}^2R_\mathrm{NS}^6\Omega_\mathrm{NS,0}^4}{6c^3}(1+t/T_\mathrm{dip}^\mathrm{iso})^{-2},
\label{eq: L_dip_iso}
\end{equation}
where $T_\mathrm{dip}^\mathrm{iso}=3c^3I_\mathrm{NS}/(B_\mathrm{NS,0}^2R_\mathrm{NS}^6\Omega_\mathrm{NS,0}^2)$ is the spin-down timescale, and $\Omega_\mathrm{NS,0}$ refers to the initial spin frequency of the magnetar. Using the same energy diffusion formula as in Equation (\ref{eq: L_rad}), the radiative luminosity of the SN powered by an isolated magnetar can be calculated as
\begin{equation}
L_\mathrm{ej,rad}^\mathrm{iso}(t)=2e^{-(t/t_\mathrm{diff})^2}\int^{t}_{0}L_\mathrm{dip}^\mathrm{iso}(1-e^{-At'^{2}})e^{(t'/t_\mathrm{diff})^2}\frac{t'\mathrm{d}t'}{t_\mathrm{diff}^2}.
\label{eq: L_rad_iso}
\end{equation}
\end{document}